# NB-IoT Small Cell: A 3GPP Perspective

Mahmoud Abbasi, *Student Member, IEEE*,

*Abstract*--Narrowband Internet of Things (NB-IoT) technology was introduced in 3GPP Release 13 to accommodate device-generated traffic over cellular networks. There have been efforts in Release 14 to deliver further improvements to facilitate the deployment of Internet of Things (IoT) use cases, such as coverage enhancement, and support for low cost, low complexity and low power consumption device. Small cell NB-IoT is added as an attractive feature in 3GPP Release 15. In this article, we provide an overview of this feature and try to shed light on major aspects of small cell deployment in NB-IoT systems..

*Index Terms*--Internet of Things, small cell, NB-IoT, IoT, 3GPP.

## I. INTRODUCTION

Machine Type Communication (MTC) has been introduced by the Third Generation Partnership Project (3GPP). MTC refers to the network of physical smart devices that are able to communicate with each other [1]. MT communication enables smart objects to play a more active role in the broad spectrum of applications, e.g. everyday life, industry, healthcare, utilities, and etc. The requirements of MTC devices in terms of data rate, delay, density, and energy can be different. For example, smart meters are stationary and need low data rate which can tolerate roughly long delay. However, all these devices use for applications that have strict requirements for device complexity (low-complexity), range (long-range), energy (low-power) and cost (low-cost).

Narrowband Internet of Things (NB-IoT) technology that introduced in 3GPP Release 13 is one category of the MTC. This technology is seen as an important step in order to accommodate IoT traffic over cellular networks [2]. NB-IoT data transmission for both Downlink (DL) and Uplink (UP) is supported in a narrow bandwidth, i.e. 180 kHz, which leads into additional 20 dB link budget in contrast to the LTE-A network. Hence, NB-IoT devices are able to work in remote and extreme coverage areas, e.g. thick-walled buildings and basements and communicate with Base Stations (BSs). For NB-IoT five new physical channels are designed due to the signals and protocols in the legacy LTE network operate at wide bandwidths, e.g. 20 MHz in LTE-A. The novel channels are as follow [3, 4]:

- Narrowband physical random access channel (NPRACH),
- Narrowband physical uplink shared channel (NPUSCH)
- Narrowband physical downlink shared channel (NPDSCH)
- Narrowband physical downlink control channel (NPDCCH)
- Narrowband physical broadcast channel (NPBCH)

Also, four novel physical signals for NB-IoT systems are defined, including narrowband reference signal (NRS), narrowband primary synchronization signal (NPSS), demodulation reference signal (DMRS) and narrowband secondary synchronization signal (NSSS). When a device decides to send data, it first has to synchronize with the BS by deriving NPSS and NSSS signals. Then, it receives NPBCH from eNB and starts NPRACH procedure. The device waits for the scheduling UL grant on NPDCCH. In the final step of NPRACH, the eNB assigns NPUSCH or NPDSCH for send and receive data to/from the BS, respectively.

One great disparity between LTE-A and NB-IoT comes from the fact that in the NB-IoT systems, IoT devices are able to spend a considerable of their lifetime in the deep sleep state and only transmit infrequent short data packets. This functionality reduces the number of required signaling messages because during the sleep state, IoT device is still registered to the network and the device needs a small number of messages to connect to the BS. Consequently, the amount of consumed energy for the connection establishment will decline dramatically [5].

Moreover, NB-IoT technology uses repetition technique for coverage enhancement, which is applicable for the reduced eNBs power. The number of repetitions depends on the device coverage level, which assigned by the BS. Please note that this technique will be discussed in the sequel.

## II. BACKGROUND TO NB-IOT ENHANCEMENTS

There have been efforts in NB-IoT/eMTC Release 14 to provide improvements mainly in two aspects. First, low cost, low complexity and low power consumption MTC device. Second, coverage enhancement in order to facilitate the deployment of IoT applications. For example, some MTC devices are deployed in the basements of buildings or foil-backed insulated locations or traditional buildings with thick walls, and these devices will suffer from significant penetration losses in contrast to LTE devices. The MTC devices in the unusual coverage areas may characterize by features such as low-to-medium data rate, not delay sensitive, and no mobility.

In Release 10, small cells deployment by low power nodes is targeted. It is expected that small cell will be able to handle the ever-growing mobile traffic, particularly in densely

Mahmoud Abbasi is with the Islamic Azad University of Mashhad, Razavi Khorasan Province, Mashhad , Iran(Mahmoud.abbasi@ieee.org).

populated indoor and outdoor areas. In general, a small cell or low-power node refers to the BS whose transmit power (Tx) is lower than the traditional macro base station. Extra functionalities are considered in small cell deployment scenarios to improve performance in hotspot indoor/outdoor locations, e.g. spectrum efficiency enhancement and improvement in discovery procedure. In the following figure, three illustrative examples of small cell deployment scenarios are provided.

In NB-IoT Release 13/14, the deployment of macro scenario for NB-IoT/eMTC applications is considered because the coverage enhancement features in the macro cells are expected to realize the requisite coverage. Furthermore, macro deployment scenario can support a large number of connections that correspond to the MTC's traffic model in the context of NB-IoT. Nevertheless, since device- and human generated communication's traffic are becoming increasingly diverse, the potential of leveraging small cell deployment scenarios, i.e. microcell, picocell, and femtocell, in NB-IoT applications has been introduced. There are many aspects with respect to the small cell support for NB-IoT, such as outdoor/indoor, spares or dense, synchronization between small cells and also between small cells and macro cell(s), and using same frequency or different frequency between small cells and macro cell (s). In addition, the same technical challenges associated with LTE small cell may true for NB-IoT small cell because of low transmit power of small cell, low coverage level, and UL/DL imbalance issue in heterogeneous network [6].

According to TS36.104, the BS classes are determined based on the minimum coupling loss between BS and User equipment (UE). The minimum coupling loss and the output power for different BS classes is provided in Table 1.

The existing cellular infrastructure should use for serving NB-IoT devices as much as possible. To achieve this goal, one key issue is how to connect the NB-IoT devices to the small cells and the macro cells in a near-optimal approach. Because of disparity between transmission power and antenna gain in the small and the macro cells, the most of NB-IoT devices will connect to the strongest DL cell, which would be the macro base station. In NB-IoT Release 15, the decoupling technique is proposed to address this issue. Later, we come back to this issue, and discuss the proposed solution in details.

## III. NB-IoT ENHANCEMENTS IN 3GPP RELEASE 15

Release 15 adds support for new features such as improved access control, small-cell support, scheduling request (SR). Furthermore, in NB-IoT Release 15, further enhancement for NB-IoT devices is introduced to improve the latency and measurement accuracy, decline power consumption, and enhance NPRACH reliability.

However, we focus on NB-IoT small cell support, and try to shed light on major aspects of deployment of microcells, picocells and femtocells in NB-IoT use cases. Since NB-IoT devices have to satisfy power saving and uplink transmitting power requirements, the main objective of NB-IoT Release 15 is the expansion in coverage for the NB-IoT devices using small cells.

### A. NB-IoT small cell architectures

In legacy LTE system, small cell has been introduced to increase the capacity of system in Heterogeneous Networks (HetNets). In this system, a dual connection architecture (DC) is designed for small cell deployment. At user plane architecture, secondary Evolved Node B (SeNB) and master eNB(MeNB) are associated with core network through S1 interface. Moreover, bearer data service separate into two parts and transmit to UE through SeNB and MeNB. In another design for user plan architecture, only MeNB is associated with core network through S1 radio interface. In the first step, the transmitted data bearer from core network transfer to MeNB and then user data send to SeNB through X2 interface. It is worth to point out that both MeNB and SeNB benefit from independent MAC entities and physical layer processing.

According to the above-described discussion and with respect to the small cell deployment scenario in legacy LTE, NB-IoT Release 15 introduces three architectures for supporting small cell in NB-IoT. In the following text, we study the proposed architectures in details.

TABLE I
Base Station Minimum Coupling Loss and Output Power

| Base station class | Minimum coupling loss | Output power |
|---|---|---|
| Wide area base station (Macro Cell) | 70 dB | There is no upper limit for this item |
| Medium range base station (Micro Cell) | 53 dB | < + 38 dBm |
| Local area base station (Pico Cell) | 45 dB | < + 24 dBm |
| Home base station (Femto Cell) | - | < + 20 dBm (in one antenna port) <br> < + 17 dBm (in two antenna ports) <br> < + 14dBm (in four antenna ports) <br> < + 11dBm (in eight antenna ports) |

#### 1) Architecture 1

Fig. 2 (a) represents Architecture 1, in which the small cells are within the coverage of the macro cell. Each small cell is associated with the core network through its own S1 interface. Moreover, independent protocol stack and complete cell feature is defined for the small cells. Here, one can also configure the small cells in a similar way to Release 14 NB-IoT cell, e.g. one anchor configured on Physical Resource Block (PRB) and multiple non-anchor PRBs. Under specific rules and policies, UEs will be able to connect to a small cell or macro cell, receive system information on NPBCH and start NPRACH procedure. Measurement, cell selection/re-selection among all the macro cells/small cells in close proximity is a part of functionalities of UEs in this architecture.

It is clear that after the deployment of small cells, it is only to be expected that the neighboring UEs would attempt to connect to the small cells. Nevertheless, from a UE viewpoint, the download coverage of small cell is generally limited compared to neighboring macro cell. Simultaneously, it is quite possible that the UE experiences better uplink coverage in small cell than macro cell. Hence, if the UE only considers Reference Signal Received Power (RSRP) as a criterion, it is difficult to select a small cell to connect to it. Three solutions, including solution 1, 2, and 3, are proposed to tackle the cell selection problem in this architecture [7]. In solution 1, the UE can use the calculated RSRP to measure path loss and then go through cell selection procedure based on the path loss value. Solution 2 proposes cell selection based on the both measured RSRP and path loss. For example, when the UE is in normal cell coverage, i.e. the calculated RSRP is equal or higher than the threshold of normal coverage, the UE need to calculate and use the path loss for cell selection. And finally, one can use different thresholds for small cell and macro cell during cell selection procedure.

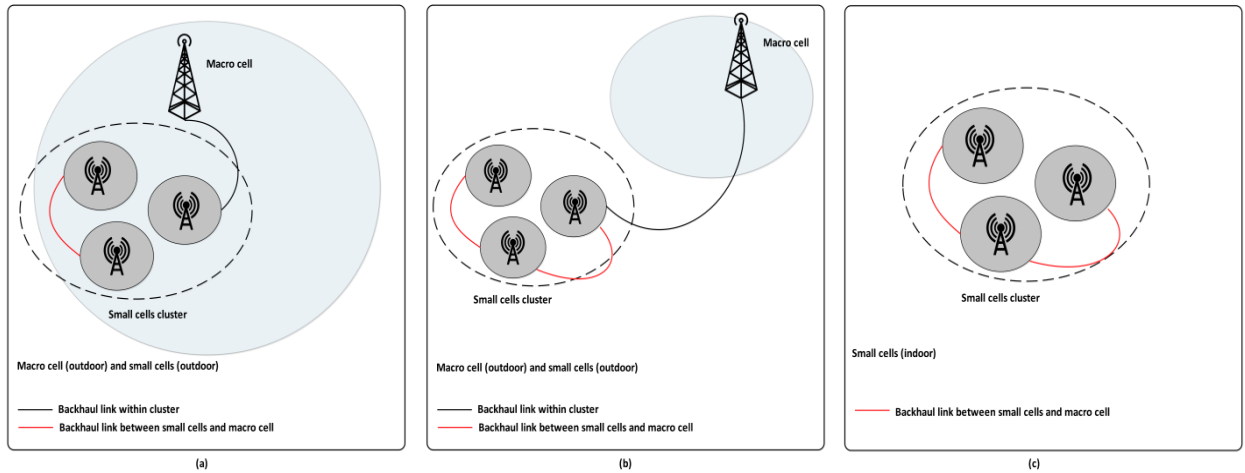

Fig. 1. Small cell scenarios: a) scenario 1; b) scenario 2; c) scenario 3.

*2) Architecture 2*

In the architecture presented in Fig. 2 (b), the macro cell is anchor eNB and can be configured with anchor PRB. The small cell will be non-anchor eNB, with non-anchor PRB. In this architecture, only anchor eNB has an S1 connection with the core network. Non-anchor eNB and anchor eNB are connected to each other through X2 link.

Such design for NB-IoT small cell architecture seems almost similar to the non-anchor PRB deployment in NB-IoT Release 14. It is reasonable to follow the non-anchor PRB mechanism in NB-IoT Release 14 as much as possible. Attention should be paid here, in NB-IoT Release 14 a small cell is configured with anchor PRB and non-anchor PRB, while in Architecture 2 only macro cell is configured with anchor PRB.

In Architecture 1, the non-anchor eNB has no Master Information Block (MIB) and System Information Block (SIB). The anchor eNB (i.e. macro cell) is responsible for broadcasting the list of the neighbor anchor eNBs and sharing the necessary information for cell selection/reselection procedure. Furthermore, when the eNB broadcasts the anchor eNBs list, the non-anchor eNBs (i.e. small cells) within the coverage of these anchor eNBs are also listed. The broadcast message may contain extra information such as frequency information, the configuration of Narrowband Reference Signal (NRS), the power of NRS, cell threshold for the non-anchor.

One must note that in this architecture when the UE decides to start RACH procedure, it may choose one eNB among the anchor eNBs and all the nearby non-anchor eNBs to find the most appropriate eNB for starting the RACH procedure. Because of the difference in small cell and macro cell coverage level, the UE should make the NRS measurement in non-anchor eNBs before it initiates RACH procedure. In this architecture, it is assumed that both anchor and non-anchor eNBs are able to transmit NRS signal. In a similar way to NB-IoT Release 14, when an eNB has been selected, the UE might choose the random access resources, e.g. PRB, subcarriers and time resource.

*3) Architecture 3*

Fig. 2 (c) represents Architecture 3, in which the macro eNB and small cells configured with the same cell identity. In this design, a small cell could also assign PRBs for non-anchor carriers. Only anchor eNB is associated with the core network through its S1 interface. Also, an X2-like connection between small cell and macro cell is considered for transmission/receiving the necessary information. Uplink/downlink transmission can take place in a separate manner and through macro cell and small cell. In such design, the primary cell is macro cell, and it is clear that all the MIB/SIB broadcasting, PRACH and paging will take place on the primary cell. Note that no broadcasted MIB/SIB would be performed on the small cells, and there is no PRACH and paging resources assignment for the non-anchor carriers of the small cell.

In order to monitor paging and start RACH procedure, the UE would connect to the macro eNB. However, during or after RACH procedure the macro eNB may reconnect the UE to the small cell to increase the efficiency of the small cell resources (e.g. through Msg4). In NB-IoT Release 13/14 a similar procedure is defined for the reconfiguration of the UE to anchor/non-anchor PRB using Msg4. In Architecture 3, there is a disparity between the small cell (i.e. non-anchor PRB) and the macro cell (i.e. anchor PRB) coverage level. Hence, before the reconnection of the UE to the small cell, the macro eNB should collect some information to determine whether the coverage level of the small cell is suitable. One option for the macro cell to assess the suitability of the small cell coverage is through X2-like interface. More specifically, the small cell can derive the macro cell's PRACH resources configuration and strive to get the preamble transmission. Then, the results of preamble reception would be delivered to the macro cell. With this approach, the UE has access to the resources of small cell for data transmission.

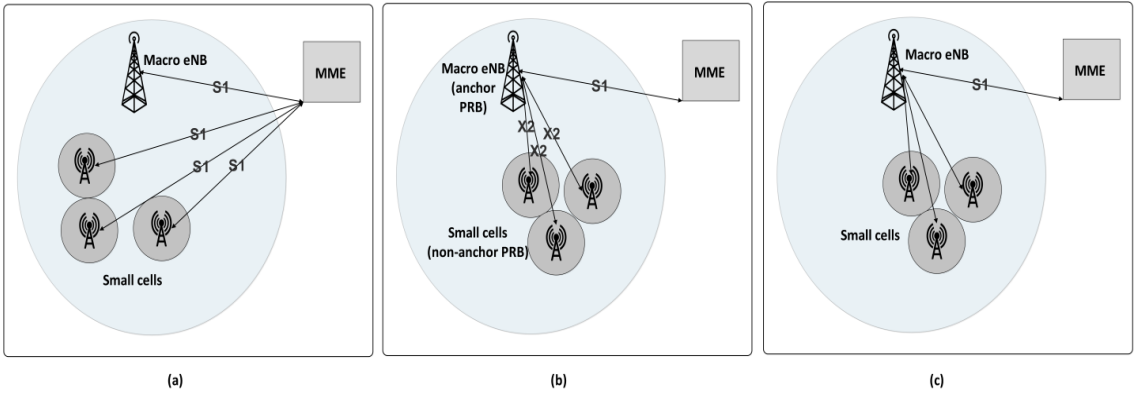

Fig. 2. NB-IoT small cell architectures: a) Architecture 1; b) Architecture 2; c) Architecture 3.

### B. Uplink aspects
#### 1) Uplink power control

Using small cell is beneficial in the improvement UL quality because of the UEs place at the close proximity of the eNBs. However, the problem in the coexistence scenario is that if a UE within a close distance to the eNB transmits at high power, it may block the receiver at the eNB for UEs in long distance. In the legacy small cell deployments, where only the open-loop power control scheme is used, e.g. for Initial Access (IA) procedure, it is crucially necessary to control the UE uplink power to avoid interference. Uplink power control controls the transmit power of the difference uplink channels, such as NPUSCH and NPRACH. However, for the current NB-IoT deployments it can be highly problematic since the coverage enhancement techniques are developed in NB-IoT systems and it is expected that the uplink repetitions are used, in which a UE would send at full power that is set by the eNB.

In open-loop power control for the UL, the UE chooses the UL to transmit power with the respect to its own measurements and parameters it gets from the network. In a HetNet NB-IoT system, including macro cells and small cells, it may necessary to establish additional rules to alleviate interference or to tackle coverage problems which are unique to the HetNet scenarios. In the following text, we focus on some HetNet scenarios where UL power control should be considered.

In the scenario depicted in Fig. 3 (a), the small-cell eNB is placed in close proximity to the macro eNB. In this deployment, UE1 is situated between macro and small-cell eNB, and camped on the small cell. This UE may introduce severe unintended interference over the macro cell uplink channels, especially when the UE is located at the edge of cell and transmit at or close to its full power. This interference can be alleviated if the UE transmits at an almost lower power. One may define the transmit power of the UE as a function of, for example, the measurement of the path loss for the strongest nearby cell by the UE.

Fig. 3 (b) shows the second scenario, in which a small cell is placed adjacent to the macro cell edge. This small cell may receive interference form the UEs that camped on the macro cell. More specifically, a macro-cell UE at the boundary of the cell may transmit at or close to its full power, and this transmission is treated as interference. To mitigate the interference, the small cell may stop to serve the UE, e.g. closed subscriber group (CSG) mode in a femtocell. In this case, in order to prevent the extra interference from such UEs, the transmit power of the users in CSG can be risen. The rise of power transmission can defined as function of, for instance, the ratio of interference-power to Johnson–Nyquist noise calculated by the femtocell.

Note that the above-mentioned interferences are not unique to NB-IoT systems but are generally common in HetNets. In the following, we investigate the UL power control issue for both NPUSCH and NPRACH.

As mentioned above, in the current NB-IoT system when a UE is in normal coverage, open-loop power control scheme is used for the uplink channels (i.e. NPRACH and NPUSCH), which means that based on the calculated path loss the UE chooses the transmit power of its UL.

As a starting point, one may refer to TS 36.213, in which the maximum UL transmission power is defined by the serving cell. Because of the coverage enhancement techniques and the uplink repetitions is introduced for NB-IoT systems, a UE may decide to transmit at full power based on what is defined by the serving cell.

In TS 36.213, the UE transmit power $P_{\text{NPUSCH},c}(i)$

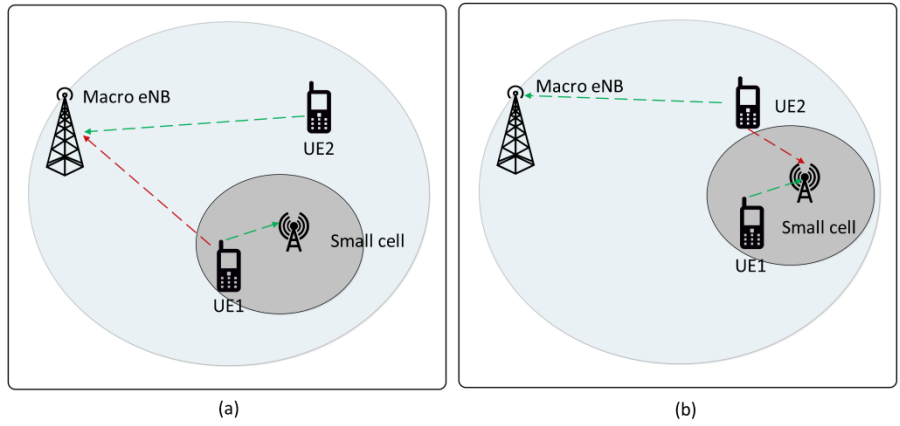

Fig. 3. Interference to macro cell from UL transmission.

for NPUSCH channel in slot i for the serving cell c is defined as follows:

If the number of repetitions for the allocated NPUSCH RUs is less than 2,

$$P_{\text{NPUSCH},c}(i) = \min \{P_{\text{CMAX},c}(i), 10 \log_{10}(M_{\text{NPUSCH},c}(i)) + P_{\text{O\_NPUSCH},c}(j) + \alpha_c(i) \cdot PL_c\} \quad (1)$$

otherwise,

$$P_{\text{NPUSCH},c}(i) = P_{\text{CMAX},c}(i) \quad (2)$$

In these expressions, $P_{\text{CMAX},c}(i)$ is the maximum allowed UE transmit power that configured by serving cell c in NB-IoT UL slot i, factor $M_{\text{NPUSCH},c}(i)$ is depend on the subcarrier spacing and its values can be {1/4, 1, 3, 6, 12}, $P_{\text{O\_NPUSCH},c}(j)$ is a parameter consist of the sum of two components provided by higher layers for serving cell $c$ where j= {1, 2}, $\alpha_c(i)$ is provided by higher layers for $j=1$, and $\alpha_c(i) = 1$ for $j= 2$. The downlink path loss for serving cell c is denoted by $PL_c$, where $PL_c$ is estimated by the UE. The interested reader may refer to [7] for a detailed description of the proposed scheme.

From above, it is seen that in a macro scenario deployment, it makes sense to set $P_{\text{CMAX},c}(i)$ as maximum transmit power that an NB-IoT UE is able to support, and when the number of repetitions for the allocated NPUSCH RUs is greater than 2, an NB-IoT UE shall transmit at its full power. These settings are proposed to achieve coverage enhancement for NB-IoT.
In small cell deployment scenario, however, in order to control the UL interference, $P_{\text{CMAX},c}(i)$ is generally set to a much smaller value to redress imbalance between the UL coverage and the DL coverage of a small cell. Consequently, the UL of the NB-IoT UE would not produce uncontrolled interference to other NB-IoT UEs in the nearby cells. Nevertheless, from the NB-IoT system perspective, this scheme is not very efficient. More specifically, one of the main reasons behind the design of NB-IoT system is to enhance the UL coverage, and hence when a NB-IoT UE is in extreme coverage and NB-IoT UE is in its maximum transmission power, further coverage enhancement is achievable by repetition. In small cell deployment, where $P_{\text{CMAX},c}(i)$ is considered smaller than the UE's maximum output power, it could be beneficial for the UE to transmit at its maximum power than $P_{\text{CMAX},c}(i)$, which is more efficient in terms of UL resource usage and UE battery life.

*2) Uplink-downlink decoupling*

As mentioned in the previous sections, small cell deployment in NB-IoT use cases can address the extended coverage issue. However, because of UL/DL imbalance in small cell, the targeting Maximum Coupling Loss (MCL) of 164 dB may no longer be achieved by the small cell. Decoupled DL/UL operation enables the NB-IoT UE to connect to the strongest eNB (macro eNB) for downlink operations and to the least path loss eNB (small-cell eNB) for uplink operations. For NB-IoT use cases which can tolerate up to 10 seconds latency this approach is a viable option because fast backhaul links between macro eNB and small-cell eNB is not essential. The decoupled UL/DL operations of NB-IoT is shown in Fig. 4, the UL of the NB-IoT UE is served by the closest small-cell eNB, and the DL is served by the strongest macro eNB. The interested reader is referred to [9, 10] for the further detailed on decoupling DL-UL operation in small cell deployment.

*C. Downlink aspects*

The eNB in small cell deployment generally has limited output power, and because up to 20 MHz bandwidth should be supported by a small-cell eNB, the actual power spectrum can be much smaller than a macro cell. Hence, the UL coverage level of small cells is usually limited. Nevertheless, a medium range BS (i.e. micro Cell) that supports NB-IoT stand-alone operation mode may provide similar coverage as a wide area BS (i.e. macro Cell) that operates in NB-IoT in-band operation mode.

In a macro cell deployment scenario, for NB-IoT in-band and guard-band mode of operation, in order to enhance the coverage, a 6 dB power boosting can be used. But, one cannot assure that the same is true for a small cell deployment. Assume that same level of power boosting applied to some type of the small-cell eNBs. In a highly dense small cell deployment scenario, since the interference problem in this situation is more severe compared to macro cell only deployment scenario, more attention should be given when we use the power boosting to keep a balance between coverage and interference management.

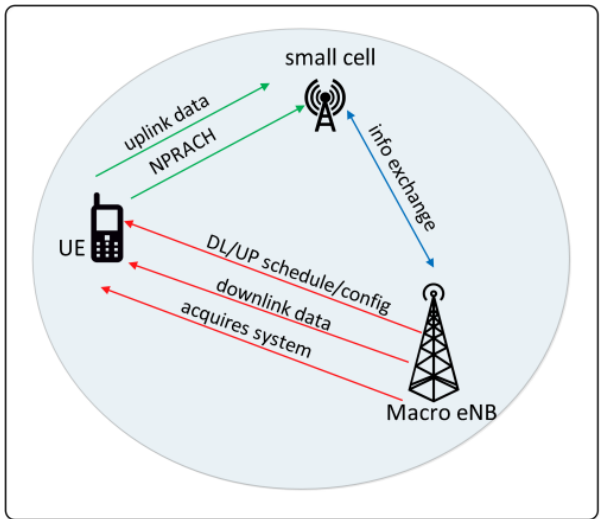

Fig. 4. Decoupled DL-UL Operation for NB-IoT.

With regard to DL power control, in NB-IoT systems, the DL power allocation is defined as a pre-determined or pre-configured power ratio between different NB-IoT physical channels/signals and no open-loop or closed-loop power control scheme for the DL is introduced. The eNB configures the DL transmit power per Resource Element (RE). For small-cell eNBs, a decrease in transmit power is considered because of the hardware characteristics of micro/pico/femto cells. Thus, the current DL power allocation strategy used in TS36.213, can be adopted for small cells.

## IV. CONCLUSTION

Providing low-complexity, long-range, low-power and low-cost connectivity over cellular networks is the key requirement for enabling networked- objects, in which every physical things that can be connected is connected. Toward this end, this paper was focused on NB-IoT 3GPP LTE Release 15, in which new features and enhancements are provided, especially small cell deployment in NB-IoT systems.

Release 15 specially focuses on the coverage enhancement using small cell. From the above-mentioned discussions, we can see that our valuable knowledge from the previous 3GPP LTE Releases paves the way for the NB-IoT systems to satisfy the requirements of NB-IoT small cell. There are many aspects with respect to NB-IoT small cell; hence the architectures and new schemes are introduced to deal with these aspects. Since small cell is an enabling technology for 5G networks and due to the peaceful coexistence of NB-IoT systems with cellular networks, NB-IoT small cell is foreseeable to be a key player in serving of IoT traffic over cellular networks.